\documentclass[aps,onecolumn,showpacs,nofootinbib,showkeys]{revtex4-2}
\usepackage[margin=3.15cm]{geometry}

\RequirePackage[T1]{fontenc}

\usepackage{epsfig,graphicx,amsmath,amssymb,bm}
\RequirePackage{mathptmx}      
\usepackage{mathrsfs}
\usepackage[english]{babel}
\usepackage{amssymb}
\RequirePackage{color}
\usepackage{changes}
\usepackage{slashed}
\usepackage{multirow}
\usepackage{scalerel}
\usepackage{tikz-feynman}
\usepackage{textcomp}
\usepackage{subcaption}
\usepackage[english]{babel}
\usepackage{float}
\RequirePackage{hyperref}
\hypersetup{
    linktocpage,
    colorlinks,
    citecolor=blue,
    filecolor=black,
    linkcolor=blue,
    urlcolor=blue,
}
\usepackage{nccmath}

\begin{document}

\title{The decays $\tau \to [K^- K^0 \pi^0, K^- K^+ \pi^-, K^0 \bar{K^0} \pi^-] \nu_{\tau}$ in the NJL quark model}


\author{M.K. Volkov$^{1}$}\email{volkov@theor.jinr.ru}
\author{A.A. Pivovarov$^{1}$}\email{tex$\_$k@mail.ru}
\author{K. Nurlan$^{1,2,3}$}\email{nurlan@theor.jinr.ru}

\affiliation{$^1$ Bogoliubov Laboratory of Theoretical Physics, JINR, 
                 141980 Dubna, Moscow region, Russia \\
                $^2$ The Institute of Nuclear Physics, Almaty, 050032, Kazakhstan \\
                $^3$ L.N. Gumilyov Eurasian National University, Astana, 010008, Kazakhstan}   


\begin{abstract}
The $\tau$ lepton decays $\tau \to [K^- K^0 \pi^0, K^- K^+ \pi^-, K ^0 \bar{K^0} \pi^-] \nu_{\tau}$ are described in the $U(3) \times U(3)$ NJL quark model. The contact channel and intermediate channels with axial-vector, vector and pseudoscalar mesons are taken into account. It is shown that the strange scalar meson $K^*_0$ plays an important role in these processes in addition to the intermediate vector meson $K^*$. The obtained results are in satisfactory agreement with the recent experimental data within the experimental and theoretical uncertainties. 


\end{abstract}


\maketitle


\section{\label{Intro}Introduction}
Numerous $\tau$ decays into mesons allow a more thoroughly study of meson interactions at low energies. One of the most effective phenomenological models for describing such interactions is the Nambu-Iona-Lasinio (NJL) quark model \cite{Nambu:1961tp,Eguchi:1976iz,Ebert:1982pk,Volkov:1984kq,Volkov:1986zb,Ebert:1985kz,Vogl:1991qt,Klevansky:1992qe,Hatsuda:1994pi,Ebert:1994mf,Bijnens:1995ww,Buballa:2003qv,Volkov:2005kw}. Using this model various $\tau$ lepton decays were successfully described \cite{Volkov:2017arr,Volkov:2022jfr}.

In recent works, three-meson $\tau$ decays $\tau \to K \pi \pi \nu_{\tau}$, $\tau \to 3K \nu_{\tau}$, $\tau \ to K \pi \eta \nu_{\tau}$, $\tau \to KK \eta \nu_{\tau}$ and $\tau \to K \eta \eta \nu_{\tau}$ \cite{ K:2023kgj,Volkov:2023evp,Volkov:2023pmy} are studied in the framework of the NJL model. However, the description of the $\tau \to K K \pi \nu_{\tau}$ process in this approach leads to a significant discrepancy theoretical results with the experiments. A more thoroughly study showed that taking into account of scalar mesons as intermediate states, which made an insignificant contribution to the description of previous $\tau$ decays, allows us to notably improve the results when describing processes $\tau \to K K \pi \nu_{\tau}$. We take into account here the contact, axial-vector, vector and pseudoscalar channels. It is shown that the axial-vector channel is dominant.

Experimental measurements of branching fractions of decays $\tau \to K K \pi \nu_{\tau}$ were carried out by the Belle, BaBar, CLEO, ALEPH, and others collaborations \cite{CLEO:1996rit,ALEPH:1997trn,ALEPH:1997cgu,ALEPH:1999jxs, CLEO:2003dfk,BaBar:2007chl,Belle:2010fal,Belle:2014mfl}. The highest statistics for these decays were obtained in recent experiments by Belle and Babar \cite{BaBar:2007chl,Belle:2010fal,Belle:2014mfl}.

\section{Effective quark-meson Lagrangian of the NJL model}
The version of the NJL model used in the present work leads to the quark-meson interaction Lagrangian of the following form \cite{Volkov:1986zb,Volkov:2005kw}
\begin{eqnarray}
	\Delta L_{int} & = & \bar{q}\left\{\sum_{i=0,\pm}\left[ig_{\pi}\gamma^{5}\lambda^{\pi}_{i}\pi^{i} +
	ig_{K}\gamma^{5}\lambda^{K}_{i}K^{i} + \frac{g_{\rho}}{2}\gamma^{\mu}\lambda^{\rho}_{i}\rho^{i}_{\mu} + \frac{g_{K^{*}}}{2}\gamma^{\mu}\lambda^{K}_{i}K^{*i}_{\mu} \right.\right. \nonumber\\
	&&\left.\left. + \frac{g_{K_1}}{2}\gamma^{\mu}\lambda^{K}_{i}K^{i}_{1A\mu}\right] + ig_{K}\gamma^{5}\lambda_{0}^{\bar{K}}\bar{K}^{0} + \frac{g_{\omega}}{2}\gamma^{\mu}\lambda^{\omega}\omega_{\mu} + \frac{g_{K^{*}}}{2}\gamma^{\mu}\lambda^{\bar{K}}_{0}\bar{K}^{*0}_{\mu}\right\}q,
\end{eqnarray}
where $q$ and $\bar{q}$ are triplets of the u, d, and s quarks with the constituent masses $m_{u} \approx m_{d} = 270$~MeV, $m_{s} = 420$~MeV, $\lambda$ are the linear combinations of the Gell-Mann matrices. 

In the given Lagrangian, the mixing of strange axial-vector mesons $K_1(1270)$ and $K_1(1400)$ are taken into account \cite{Volkov:2022jfr, Volkov:1984gqw,Suzuki:1993yc}:
\begin{eqnarray}
    K_{1A} & = & K_1(1270)\sin{\alpha} + K_1(1400)\cos{\alpha}, \nonumber\\
    K_{1B} & = & K_1(1270)\cos{\alpha} - K_1(1400)\sin{\alpha},
\end{eqnarray}
where $\alpha = 57^\circ$. 
The $K_{1B}$ state is not taken into account by the NJL model. The contribution from these state is proportional to the effects of $U(3)$ symmetry breaking and can be taken into account in the model uncertainty \cite{Volkov:2022jfr}.

The quark-meson coupling constants appear from the renormalization of the Lagrangian:
\begin{displaymath}
	g_{\pi} = \sqrt{\frac{Z_{\pi}}{4 I_{20}}}, \quad g_{\rho} = g_{\omega} = \sqrt{\frac{3}{2 I_{20}}}, \quad g_{K} = \sqrt{\frac{Z_{K}}{4 I_{11}}}, \quad g_{K^{*}} = g_{K_1} = \sqrt{\frac{3}{2 I_{11}}},
\end{displaymath}
where
\begin{eqnarray}
	&Z_{\pi} = \left(1 - 6\frac{m^{2}_{u}}{M^{2}_{a_{1}}}\right)^{-1}, \quad
	Z_{K} = \left(1 - \frac{3}{2}\frac{(m_{u} + m_{s})^{2}}{M^{2}_{K_{1A}}}\right)^{-1},& \nonumber\\
	&M^{2}_{K_{1A}} = \left(\frac{\sin^{2}{\alpha}}{M^{2}_{K_{1}(1270)}} + \frac{\cos^{2}{\alpha}}{M^{2}_{K_{1}(1400)}}\right)^{-1},&
\end{eqnarray}
$Z_{\pi}$ and $Z_{K}$ are the factors describing $\pi - a_1$, $\eta - f_{1}$ and $K - K_{1}$ transitions, $M_{a_{1}} = 1230$~MeV, $M_{f_{1}} = 1426$~MeV, $M_{K_{1}(1270)} = 1253$~MeV, $M_{K_{1}(1400)} = 1403$~MeV \cite{ParticleDataGroup:2022pth} are the masses of the axial vector mesons $a_{1}$ and $K_{1}$.

The integrals in the definitions of the coupling constants take the form:
\begin{equation}
\label{integral}
	I_{nm} = -i\frac{N_{c}}{(2\pi)^{4}}\int\frac{\theta(\Lambda^{2} + k^2)}{(m_{u}^{2} - k^2)^{n}(m_{s}^{2} - k^2)^{m}}
	\mathrm{d}^{4}k,
\end{equation}
where $\Lambda = 1265$~MeV is the cut-off parameter~\cite{Volkov:2022jfr}.

\section{The $\tau \to K^-K^0\pi^0\nu_{\tau}$ decay amplitude}
\begin{figure*}[t]
 \centering
  \begin{subfigure}{0.5\textwidth}
   \centering
    \begin{tikzpicture}
     \begin{feynman}
      \vertex (a) {\(\tau^-\)};
      \vertex [dot, right=2cm of a] (b){};
      \vertex [above right=2cm of b] (c) {\(\nu_{\tau}\)};
      \vertex [dot, below right=1.2cm of b] (d) {};
      \vertex [dot, above right=1.2cm of d] (e) {};
      \vertex [dot, below right=1.2cm of d] (h) {};
      \vertex [dot, right=1.6 cm of e] (f) {};
      \vertex [dot, above right=1.0cm of f] (n) {};  
      \vertex [dot, below right=1.0cm of f] (m) {};   
      \vertex [right=1.2cm of n] (l) {\(\ K \)}; 
      \vertex [right=1.2cm of m] (s) {\(\pi (K) \)};  
      \vertex [right=1.4cm of h] (k) {\(K (\pi) \)}; 
      \diagram* {
         (a) -- [fermion] (b),
         (b) -- [fermion] (c),
         (b) -- [boson, edge label'=\(W^-\)] (d),
         (d) -- [fermion] (e),  
         (e) -- [fermion] (h),
         (d) -- [anti fermion] (h),
         (e) -- [edge label'=\({ K^*, K^*_0 (\rho)} \)] (f),
         (f) -- [fermion] (n),
         (n) -- [fermion] (m),
         (f) -- [anti fermion] (m), 
         (h) -- [] (k),
         (n) -- [] (l),
	 (m) -- [] (s),
      };
     \end{feynman}
    \end{tikzpicture}
  \end{subfigure}%
 \caption{The contact diagram of the decay$\tau \to K K \pi  \nu_\tau$.}
 \label{diagram1}
\end{figure*}
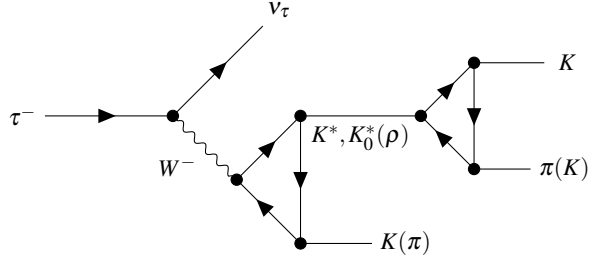

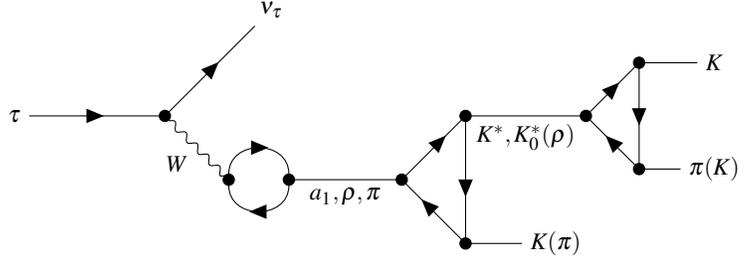
\begin{figure*}[t]
 \centering
 \centering
 \begin{subfigure}{0.5\textwidth}
  \centering
   \begin{tikzpicture}
    \begin{feynman}
      \vertex (a) {\(\tau\)};
      \vertex [dot, right=2cm of a] (b){};
      \vertex [above right=2cm of b] (c) {\(\nu_{\tau}\)};
      \vertex [dot, below right=1.2cm of b] (d) {};
      \vertex [dot, right=0.8cm of d] (l) {};
      \vertex [dot, right=1.5cm of l] (g) {};
      \vertex [dot, above right=1.2cm of g] (e) {};
      \vertex [dot, below right=1.2cm of g] (h) {};      
      \vertex [dot, right=1.6cm of e] (f) {};
      \vertex [dot, above right=1.0cm of f] (n) {};
      \vertex [dot, below right=1.0cm of f] (m) {};
      \vertex [right=1.0cm of n] (s) {\( K \)};
      \vertex [right=1.0cm of m] (r) {\( \pi (K) \)};
      \vertex [right=1.2cm of h] (k) {\( K (\pi) \)}; 
      \diagram* {
         (a) -- [fermion] (b),
         (b) -- [fermion] (c),
         (b) -- [boson, edge label'=\(W\)] (d),
         (d) -- [fermion, inner sep=1pt, half left] (l),
         (l) -- [fermion, inner sep=1pt, half left] (d),
         (l) -- [edge label'=\({ a_1, \rho, \pi} \)] (g),
         (g) -- [anti fermion] (h),  
         (h) -- [anti fermion] (e),
         (e) -- [anti fermion] (g),      
         (e) -- [edge label'=\( {K^*, K^*_0 (\rho)} \)] (f),
         (f) -- [fermion] (n),
         (n) -- [fermion] (m),
         (m) -- [fermion] (f),
         (h) -- [] (k),
         (n) -- [] (s),
         (m) -- [] (r),
      };
     \end{feynman}
    \end{tikzpicture}
  \end{subfigure}%
 \caption{The diagram with the intermediate mesons describing the decay $\tau \to K K \pi \nu_\tau$.}
 \label{diagram2}
\end{figure*}

Diagrams describing the decay of $\tau \to K^-K^0\pi^0\nu_{\tau}$ are presented in Figs. \ref{diagram1} and \ref{diagram2}.

The amplitude of this decay, depending on the type of the first intermediate resonance, including axial-vector, vector and pseudoscalar channels, takes the following form:
\begin{eqnarray}
\label{amplitude}
    \mathcal{M} = G_{F} V_{ud} F_{\pi} g_{\rho}^2 L_{\mu} \left\{\mathcal{M}_{A} + \mathcal{M}_{V} + \mathcal{M}_{P}\right\}^{\mu},
\end{eqnarray}
where $G_{F}$ is the Fermi constant, $V_{us}$ is the element of the Cabibbo-Kobayashi-Maskawa matrix, $L_{\mu}$ is the weak lepton current.
Each channel includes a sum of subprocesses in which the second intermediate resonances are vector or scalar mesons:
\begin{eqnarray}
\label{contributions}
    \mathcal{M}_{A}^{\mu} & = & -i\frac{\sqrt{2}}{2} Z_{\pi} BW_{a_1}^q \left[g^{\mu\nu}h_{a_1} - q^{\mu}q^{\nu}\right] \nonumber\\
    && \times\left\{\left(p_{K^0} - p_{K^-}\right)_{\nu}BW_{\rho^-}^s + \frac{Z_K}{4} \left(3 - \frac{m_s}{m_u}\right)\left(\frac{p_{\pi}}{Z_{a_{1}}} - \frac{p_{K^-}}{Z_{K_{1}}}\right)_{\nu}BW_{K^{*-}}^u - \frac{Z_K}{4} \left(3 - \frac{m_s}{m_u}\right)\left(\frac{p_{\pi}}{Z_{a_{1}}} - \frac{p_{K^0}}{Z_{K_{1}}}\right)_{\nu}BW_{K^{*0}}^t \right. \nonumber\\
    && \left. + \frac{Z_K}{3} \frac{m_s}{m_u}\left[\left(p_{K^0} - p_{K^-} - p_{\pi}\right)_{\nu}BW_{K_{0}^{*-}}^u - \left(p_{K^-} - p_{K^0} - p_{\pi}\right)_{\nu}BW_{K_{0}^{*0}}^t\right]\right\}, \nonumber\\
    \mathcal{M}_{V}^{\mu} & = & \sqrt{2} g_{K}^2 Z_{\pi} \left(\frac{1}{Z_{a_{1}}} + \frac{1}{Z_{K_{1}}}\right) I_{c} h_{\rho} BW_{\rho}^q \left(BW_{K^{*-}}^u - BW_{K^{*0}}^t \right)  e^{\mu \nu \lambda \delta} p_{\pi \nu} p_{K^{0}\lambda} p_{K^{-}\delta}, \nonumber\\
    \mathcal{M}_{P}^{\mu} & = & -i \sqrt{2} BW_{\pi}^q q^{\mu} \left\{p_{\pi}^\nu \left(p_{K^0} - p_{K^-}\right)_{\nu}BW_{\rho^-}^s \right. \nonumber\\
    && \left.+ \frac{Z_\pi Z_K}{4} \left(\frac{q}{Z_{a_{1}}} + \frac{p_{K^0}}{Z_{K_{1}}}\right)^{\nu} \left(\frac{p_{\pi}}{Z_{a_{1}}} - \frac{p_{K^-}}{Z_{K_{1}}}\right)_{\nu} BW_{K^{*-}}^u - \frac{Z_\pi Z_K}{4} \left(\frac{q}{Z_{a_{1}}} + \frac{p_{K^-}}{Z_{K_{1}}}\right)^{\nu} \left(\frac{p_{\pi}}{Z_{a_{1}}} - \frac{p_{K^0}}{Z_{K_{1}}}\right)_{\nu} BW_{K^{*0}}^t \right. \nonumber\\
    && \left. + \frac{2}{3}Z_\pi Z_K m_s^2 \left(BW_{K^{*0}_0}^t - BW_{K^{*-}_0}^u\right)\right\},
\end{eqnarray}
where $p_{\pi}$, $p_{K^-}$ and $p_{K^0}$ are the momenta of final mesons, $q$ is the momentum of the first intermediate meson.

The factors $Z_{K_1}$ and $Z_{a_1}$ appear as a result of the explicit allowance for transitions between the axial vector and pseudoscalar states in the different diagram vertices:
\begin{eqnarray}
    Z_{a_{1}} = \left(1 - 3\frac{m_{u}(3m_{u} - m_{s})}{M_{a_{1}}^{2}}\right)^{-1}, \quad Z_{K_{1}} = \left(1 - 3\frac{m_{s}(m_{u} + m_{s})}{M_{K_{1A}}^{2}}\right)^{-1}.
\end{eqnarray}
The factors $h_{a_1}$ and $h_{\rho}$ reads
\begin{eqnarray}
    h_{a_1} & = & M^2_{a_1} - 6m^2_u - i\sqrt{q^{2}}\Gamma_{a_1}, \\
    h_{\rho} & = & M^2_{\rho} - i\sqrt{q^{2}}\Gamma_{\rho}.
\end{eqnarray}

The intermediate states are discribed using the Breit-Wigner propagator:
\begin{eqnarray}
    BW^q_{a_1} = \frac{1}{M_{a_1}^{2} - q^{2} - i\sqrt{q^2} \, \Gamma_{a_1}}, \nonumber\\   
    BW^s_{\rho^-} = \frac{1}{M_{\rho^-}^{2} - s - i\sqrt{s} \, \Gamma_{\rho^-}}, \nonumber\\       
    BW^u_{K^{*-}} = \frac{1}{M_{K^{*-}}^{2} - u - i\sqrt{u} \, \Gamma_{K^{*-}}}, \nonumber\\ 
    BW^t_{K^{*0}} = \frac{1}{M_{K^{*0}}^{2} - t - i\sqrt{t} \,
    \Gamma_{K^{*0}}},
\end{eqnarray}
where $q=p_{K^-} + p_{K^0} + \pi^0$, $u = {(p_{K^-} + p_{\pi^0})}^2$, $t = {(p_{K^0} + p_{\pi^0})}^2$, $s = {(p_{K^0} + p_{K^-})}^2$. The mass and widths of mesons are taken from PDG \cite{ParticleDataGroup:2022pth}.

In the vector channel, the combinations of the convergent integrals appear:
\begin{eqnarray}
    I_c = m_{s} I_{21} - (m_{s} - m_{u}) \left( I_{21} + m^2_u I_{31}\right).
\end{eqnarray}

The total branching fractions of this process is:
\begin{eqnarray}
    Br(\tau \to K^-K^0\pi^0\nu_{\tau}) & = & 1.48 \times 10^{-3}.
\end{eqnarray}

Without taking into account scalar meson states, the total branching fractions takes the value
\begin{eqnarray}
    Br(\tau \to K^-K^0\pi^0\nu_{\tau}) & = & 0.52 \times 10^{-3}.
\end{eqnarray}

The contribution from the axial-vector channel for this process is
\begin{eqnarray}
    Br(\tau \to K^-K^0\pi^0\nu_{\tau})_{A} & = & 1.46 \times 10^{-3}.
\end{eqnarray}

Without taking into account scalar meson states, the same channel gives
\begin{eqnarray}
    Br(\tau \to K^-K^0\pi^0\nu_{\tau})_{A} & = & 0.49 \times 10^{-3}.
\end{eqnarray}

The contribution of the pseudoscalar channel for this process is
\begin{eqnarray}
    Br(\tau \to K^-K^0\pi^0\nu_{\tau})_{P} & = & 0.025 \times 10^{-3}.
\end{eqnarray}

Without taking into account scalar meson states, the pseudoscalar channel gives
\begin{eqnarray}
    Br(\tau \to K^-K^0\pi^0\nu_{\tau})_{P} & = & 0.024 \times 10^{-3}.
\end{eqnarray}

The contribution of the vector channel is
\begin{eqnarray}
    Br(\tau \to K^-K^0\pi^0\nu_{\tau})_{V} & = & 0.038 \times 10^{-3}.
\end{eqnarray}   

\section{The $\tau \to K^-K^+\pi^-\nu_{\tau}$ decay amplitude}
The decay of $\tau \to K^- K^+ \pi^- \nu_{\tau}$, unlike previous decay, proceeds through intermediate channels with the neutral second resonances $K^{*0}$ or $K^{*0}_0$. 
The contributions from the axial-vector, vector and pseudoscalar channels take the form:
\begin{eqnarray}
\label{contributions2}
    \mathcal{M}_{A}^{\mu} & = & \frac{i}{2} Z_{\pi} BW_{a_1}^q \left[g^{\mu\nu}h_{A} - q^{\mu}q^{\nu}\right] \nonumber\\
    && \times\left\{\left(p_{K^+} - p_{K^-}\right)_{\nu}BW_{\rho^0}^s + \frac{Z_K}{2} \left(3 - \frac{m_s}{m_u}\right)\left(\frac{p_{K^+}}{Z_{K_{1}}} - \frac{p_{\pi}}{Z_{a_{1}}}\right)_{\nu}BW_{K^{*0}}^t\right. \nonumber\\
    && \left. + \frac{2}{3} Z_K \frac{m_s}{m_u}
    \left(p_{K^+} + p_{\pi^-} - p_{K^-}\right)_{\nu}BW_{K_{0}^{*0}}^t + \frac{2}{3} Z_K \left(2 - \frac{m_s}{m_u}\right)\left(p_{K^+} + p_{K^-} - p_{\pi^-}\right)_{\nu}BW_{f_{0}}^s\right\}, \nonumber\\
    \mathcal{M}_{V}^{\mu} & = &  2 h_{V} BW_{\rho}^q\left\{2 g_{\pi}^2 I_{30} BW_{\omega}^s + g_{K}^2 Z_{\pi} \left(\frac{1}{Z_{a_{1}}} + \frac{1}{Z_{K_{1}}}\right) I_{c} BW_{K^{*0}}^t \right\} e^{\mu \nu \lambda \delta} p_{\pi^- \nu} p_{K^{+}\lambda} p_{K^{-}\delta}, \nonumber\\
    \mathcal{M}_{P}^{\mu} & = & i BW_{\pi}^q q^{\mu} \left\{p_{\pi^+}^\nu \left(p_{K^+} - p_{K^-}\right)_{\nu}BW_{\rho^0}^s + \frac{Z_\pi Z_K}{2} \left(\frac{q}{Z_{a_{1}}} + \frac{p_{K^-}}{Z_{K_{1}}}\right)^{\nu} \left(\frac{p_{K^+}}{Z_{K_{1}}} - \frac{p_{\pi^-}}{Z_{a_{1}}}\right)_{\nu} BW_{K^{*0}}^t \right. \nonumber\\
    && \left. + \frac{4}{3}Z_\pi Z_K m_u^2 \left(2 - \frac{m_s}{m_u}\right) BW_{f_{0}}^s + \frac{4}{3}Z_\pi Z_K m_s^2 BW_{K^{*0}_0}^t \right\},
\end{eqnarray}
where $p_{K^{-}}$, $p_{K^{+}}$ and $p_{\pi^-}$ are the momenta of the final mesons.

In this decay, the contribution from the strange scalar meson $K^*_0$ as a second resonance also plays a significant role. Its separate contribution in the axial-vector channel to the branching fractions is $Br(\tau \to K^- K^+ \pi^- \nu_\tau) = 0.4 \times 10^{-3}$.  

The decay amplitude of $\tau \to \pi^- K^0 \bar{K^0} \nu_\tau$ is obtained from (\ref{contributions2}) with the replacement of the masses of neutral kaons in the final state and in the second intermediate resonance, instead of neutral $K^{*0}$ and $K^{*0}_0$ there will be charged mesons $K^{*-}$ and $K^{*-}_0$.

The calculated branching fractions for all considered decays within the NJL model and comparison with experimental data are given in Table \ref{Tab1}.

\begin{table}[h!]
\begin{center}
\begin{tabular}{cccc}
\hline
   & Br$(\tau \to K^- K^0 \pi^0 \nu_{\tau})$ & Br$(\tau \to K^0 \bar{K^0}\pi^-\nu_{\tau})$ & Br$(\tau \to K^- K^+ \pi^-\nu_{\tau})$ \\
\hline
NJL model & $1.48$ & $0.95$ & $1.04$ \\
\hline
PDG \cite{ParticleDataGroup:2022pth} & $1.50 \pm 0.07$ & $1.55 \pm 0.24$ & $1.43 \pm 0.02$\\
BaBar & -- & -- & $1.34 \pm 0.03$ \cite{BaBar:2007chl} \\
Belle & $1.49 \pm 0.07$ \cite{Belle:2014mfl} & -- & $1.55 \pm 0.01$ \cite{Belle:2010fal} \\
\hline
\end{tabular}
\end{center}
\caption{The comparison of the branching fractions for the decays $\tau \to K K \pi \nu_{\tau}$ (in $Br \times 10^{-3}$).}
\label{Tab1}
\end{table}

\subsection*{Acknowledgments}
The authors thank A.A. Osipov for useful discussions.  

\section{Conclusion}
Recently in the framework of the $U(3) \times U(3)$ quark NJL model, three-meson tau-lepton decays $\tau \to K \pi \pi \nu_{\tau}$, $\tau \to 3K \nu_{\tau}$, $\tau \to K \pi \eta \nu_{\tau}$, $\tau \to KK \eta \nu_{\tau}$ and $\tau \to K \ eta \eta \nu_{\tau}$ are described in quite satisfactory agreement with experimental data \cite{K:2023kgj,Volkov:2023evp,Volkov:2023pmy}.
The present paper is a natural conclusion of recent work on the description of three-meson tau decays. As in previous decays, the dominant role in this process is played by the intermediate axial-vector channel. However, unlike previous decays, in the present decay $\tau \to K K \pi \nu_{\tau}$, taking into account the contribution from the intermediate scalar meson $K^*_0$ plays a very important role.

Indeed, considering the contribution in the axial-vector channel only from the additional vector resonance leads to a very small value of the branching fraction of decay $Br(\tau \to K^-K^0\pi^0\nu_{\tau}) = 0.52 \times 10^{-3}$. A comparable contribution is given by taking into account the scalar intermediate resonance.
In the case of previous decays, where the main intermediate axial-vector meson was the strange $K_1$, the contribution from the additional scalar meson was negligible. In this way, the decays considered here differ markedly from the above processes.      

A similar description of the decay of $\tau \to K K \pi \nu_{\tau}$ was made in the work \cite{Dumm:2009kj} within the resonance chiral theory (R$\chi$T). However, in this work the mass of the intermediate axial vector meson $M_{a_1}=1120$ MeV deviates noticeably from the standard value given in the PDG $M_{a_1} = 1230 \pm 40$ MeV \cite{ParticleDataGroup:2022pth}. In our case, this deviation is much less than $M_{a_1}=1270$ MeV instead of $M_{a_1}=1230$ MeV. In addition, the work \cite{Dumm:2009kj} notes the dominant role of the vector channel. At the same time, in all the processes we described, the axial-vector channel is dominant.


\begin{thebibliography}{99}
\bibitem{Nambu:1961tp}
Y.~Nambu and G.~Jona-Lasinio,
Phys. Rev. \textbf{122} (1961), 345-358

\bibitem{Eguchi:1976iz} 
  T.~Eguchi,
  Phys.\ Rev.\ D {\bf 14}, 2755 (1976).

\bibitem{Ebert:1982pk}
   D.~Ebert and M.~K.~Volkov,
    Z. Phys. C \textbf{16} (1983), 205

\bibitem{Volkov:1984kq}
    M.~K.~Volkov,
    Annals Phys. \textbf{157} (1984), 282-303

\bibitem{Volkov:1986zb} 
  M.~K.~Volkov,
  Sov.\ J.\ Part.\ Nucl.\  {\bf 17}, 186 (1986)
  [Fiz.\ Elem.\ Chast.\ Atom.\ Yadra {\bf 17}, 433 (1986)].

\bibitem{Ebert:1985kz} 
  D.~Ebert and H.~Reinhardt,
  Nucl.\ Phys.\ B {\bf 271}, 188 (1986).

\bibitem{Vogl:1991qt} 
  U.~Vogl and W.~Weise,
  Prog.\ Part.\ Nucl.\ Phys.\  {\bf 27}, 195 (1991).
  
\bibitem{Klevansky:1992qe} 
  S.~P.~Klevansky,
  Rev.\ Mod.\ Phys.\  {\bf 64}, 649 (1992).

\bibitem{Hatsuda:1994pi} 
  T.~Hatsuda and T.~Kunihiro,
  Phys.\ Rept.\  {\bf 247}, 221 (1994).
  
\bibitem{Ebert:1994mf} 
  D.~Ebert, H.~Reinhardt and M.~K.~Volkov,
  Prog.\ Part.\ Nucl.\ Phys.\  {\bf 33}, 1 (1994).

\bibitem{Bijnens:1995ww}
J.~Bijnens,
Phys. Rept. \textbf{265} (1996), 369-446

\bibitem{Buballa:2003qv}
    M.~Buballa,
    Phys. Rept. \textbf{407}, 205-376 (2005)

\bibitem{Volkov:2005kw} 
  M.~K.~Volkov and A.~E.~Radzhabov,
  Phys.\ Usp.\  {\bf 49}, 551 (2006).

\bibitem{Volkov:2017arr}
M.~K.~Volkov and A.~B.~Arbuzov,
Phys. Usp. \textbf{60} (2017) no.7, 643-666

\bibitem{Volkov:2022jfr}
   M.~K.~Volkov, A.~A.~Pivovarov and K.~Nurlan,
    Symmetry \textbf{14} (2022) no.2, 308.

\bibitem{K:2023kgj}
M.~K. Volkov, A.~A. Pivovarov and K.~Nurlan,
Phys. Rev. D \textbf{107} (2023) no.11, 116009

\bibitem{Volkov:2023evp}
M.~K.~Volkov, A.~A.~Pivovarov and K.~Nurlan,
Eur. Phys. J. A \textbf{59} (2023) no.7, 175

\bibitem{Volkov:2023pmy}
M.~K.~Volkov, A.~A.~Pivovarov and K.~Nurlan,
[arXiv:2307.09228 [hep-ph]].

\bibitem{CLEO:1996rit}
T.~E.~Coan \textit{et al.} [CLEO],
Phys. Rev. D \textbf{53} (1996), 6037-6053

\bibitem{ALEPH:1997trn}
R.~Barate \textit{et al.} [ALEPH],
Eur. Phys. J. C \textbf{4} (1998), 29-45

\bibitem{ALEPH:1997cgu}
R.~Barate \textit{et al.} [ALEPH],
Eur. Phys. J. C \textbf{1} (1998), 65-79

\bibitem{ALEPH:1999jxs}
R.~Barate \textit{et al.} [ALEPH],
Eur. Phys. J. C \textbf{10} (1999), 1-18

\bibitem{CLEO:2003dfk}
R.~A.~Briere \textit{et al.} [CLEO],
Phys. Rev. Lett. \textbf{90} (2003), 181802

\bibitem{BaBar:2007chl}
B.~Aubert \textit{et al.} [BaBar],
Phys. Rev. Lett. \textbf{100} (2008), 011801

\bibitem{Belle:2010fal}
M.~J.~Lee \textit{et al.} [Belle],
Phys. Rev. D \textbf{81} (2010), 113007

\bibitem{Belle:2014mfl}
S.~Ryu \textit{et al.} [Belle],
Phys. Rev. D \textbf{89} (2014) no.7, 072009

\bibitem{Volkov:1984gqw}
    M.~K.~Volkov and A.~A.~Osipov,
    Sov. J. Nucl. Phys. \textbf{41}, 500-503 (1985).

\bibitem{Suzuki:1993yc}
    M.~Suzuki,
    Phys. Rev. D \textbf{47} (1993), 1252-1255.

\bibitem{ParticleDataGroup:2022pth}
R.~L.~Workman \textit{et al.} [Particle Data Group],
PTEP \textbf{2022} (2022), 083C01

\bibitem{Dumm:2009kj}
D.~G.~Dumm, P.~Roig, A.~Pich and J.~Portoles,
Phys. Rev. D \textbf{81} (2010), 034031

\end{thebibliography}
\end{document}